\documentclass[11pt]{article}

\usepackage{latexsym,amsmath,amssymb}
\usepackage{epsf}
\usepackage{graphicx}

\newcommand{\lsim}{\lesssim}
\newcommand{\gsim}{\gtrsim}

\newcommand{\br}{\mathop{\mathrm{Br}}}

\newcommand{\beq}{\begin{equation}}   
\newcommand{\eeq}{\end{equation}}
\newcommand{\bea}{\begin{eqnarray}}   
\newcommand{\eea}{\end{eqnarray}}
\def\GEV#1{10^{#1}{\rm\,GeV}}

\def\gev{{\rm\,GeV}}

\def\lrf#1#2{ \left(\frac{#1}{#2}\right)}
\def\lrfp#1#2#3{ \left(\frac{#1}{#2} \right)^{#3}}
\def\vev#1{\langle #1\rangle}

\let\bar\overline \let\tilde\widetilde

\def\s#1{_\mathrm{#1}}             
\def\diff#1#2{\frac{d{#1}}{d{#2}}} 

\makeatletter
\def\EE{\@ifnextchar-{\@@EE}{\@EE}}                        
\def\@EE#1{\ifnum#1=1 \times\!10 \else \times\!10^{#1}\fi} 
\def\@@EE#1#2{\times\!10^{-#2}}                            
\makeatother

\newcommand{\dP}[1][p]{\!\frac{d^3#1}{(2\pi)^3}} 

 \def\muL{{\mu\s L}} 
 \def\muR{{\mu\s R}} \def\tauR{{\tau\s R}}
\def\numu{{\nu_\mu}}

\def\Hd{{H_d}} 

\def\bigvev#1{\Bigl\langle#1\Bigr\rangle}
\def\effvev#1{\Bigl\langle\!\!\Big\langle#1\Big\rangle\!\!\Bigr\rangle}
\def\funcT#1#2{\,F_{#1}\!\left(\frac{#2}{T}\right)}
\def\funcTs#1#2{\,F_{#1}\!\left(#2/T\right)}
\def\restrict#1#2{\left.#1\right|_{#2}}

\def\eqref#1{\textup{Eq.~(\ref{#1})}}

\def\squarkmass{m_{\tilde q}}
\def\sleptonmass{m_{\tilde\ell}}
\def\higgsinomass{m_{\tilde H}}

\def\UDD{\bar U\,\bar D\,\bar D}

\def\geff{\,g_{\rm eff}}
\def\funcgeff#1{\,g_{\rm eff}\!\left(\frac{#1}{T}\right)}

\begin{document}
\baselineskip=18pt

\begin{titlepage}

\begin{flushright}
UT--09--27\\
IPMU--09--0145\\
\end{flushright}

\vskip 1.35cm
\begin{center}
{\Large \bf
Lepton Flavor Violation and\\
Cosmological Constraints on R-parity Violation
}
\vskip 1.2cm
Motoi Endo$^{1,2}$, Koichi Hamaguchi$^{1,2}$, Sho Iwamoto$^{1}$
\vskip 0.4cm

{\it $^1$  Department of Physics, University of Tokyo,
Tokyo 113-0033, Japan\\
$^2$ Institute for the Physics and Mathematics of the Universe, 
University of Tokyo,\\ Chiba 277-8568, Japan
}

\vskip 1.5cm

\abstract{ 
In supersymmetric standard models R-parity violating couplings 
are severely constrained, since otherwise they would erase 
the existing baryon asymmetry before the electroweak transition.
It is often claimed that this cosmological constraint 
can be circumvented if the baryon number and 
one of the lepton flavor numbers are sufficiently
conserved in these R-parity violating couplings, because $B/3-L_i$
for each lepton flavor is separately conserved by the sphaleron
process.
We discuss the effect of lepton flavor violation on the $B-L$ conservation,
and show that even tiny slepton mixing angles 
$\theta_{12}\gsim {\cal O}(10^{-4})$ 
and $\theta_{23}, \theta_{13}\gsim {\cal O}(10^{-5})$ will spoil the 
separate $B/3-L_i$ conservation.  
In particular, if lepton flavor violations are observed 
in experiments such as MEG and B-factories, it will imply that 
{\it all} the R-parity violating couplings must be suppressed
to avoid the $B-L$ erasure. 
We also discuss the implication for the decay of the 
lightest MSSM particle at the LHC.
}
\end{center}
\end{titlepage}

\setcounter{footnote}{0}
\setcounter{page}{2}

\section{Introduction}

In most of supersymmetric (SUSY) standard models, R-parity
is assumed to be exactly conserved, which prohibits the following baryon and
lepton number violating operators,
\beq W_{\rm RpV} = 
\frac{1}{2}\lambda_{ijk}L_i L_j \bar{E}_k 
+ \lambda'_{ijk} L_i Q_j \bar{D}_k 
+ \frac{1}{2}\lambda''_{ijk} \bar{U}_i\bar{D}_j\bar{D}_k 
+ \mu_i L_i H_u\,.  
\eeq
R-parity guarantees the stability of
the lightest SUSY particle (LSP) and plays a crucial role in LHC
physics.  In R-parity conserving models with the neutralino LSP,
all the SUSY signals include a missing transverse momentum.
If gravitino is the LSP, R-parity conservation
leads to a long lifetime of the next-to-LSP (NLSP), which becomes a
stable particle in collider scale.
In fact, most of the LHC studies of SUSY signals crucially rely on these
features.

However, if R-parity is broken, the LSP is no longer stable, 
and the LHC signatures of SUSY events may
drastically change. In particular, the decay length and the decay mode of
the lightest ordinary SUSY particle 
(LOSP)\footnote{LOSP denotes the lightest SUSY particle among the superpartners of
the standard model particles. If gravitino is the LSP, NLSP is the LOSP. 
If not, LSP is the LOSP.}
crucially depend on the pattern and the size of the R-parity
violating couplings~\cite{Barbier:2004ez}. 
It is therefore very important to know what are the allowed R-parity violating couplings.

The most stringent constraint on R-parity violating couplings
comes from cosmology.  Assuming that the baryon asymmetry is generated
before the electroweak phase transition, baryon
or lepton number violating processes induced by the R-parity breaking
couplings (together with the sphaleron process~\cite{Kuzmin:1985mm})
would wash out the existing baryon asymmetry unless these couplings are
sufficiently suppressed. The bound is roughly given by~\cite{RpVcosmobound}
\beq
\lambda, \lambda', \lambda'' \lsim {\cal O}(10^{-7})\,.
\label{eq:NaiveCosmoBound}
\eeq
The bounds from the laboratory experiments and neutrino masses
are much weaker than this cosmological bound
(cf.~\cite{Barbier:2004ez,RpV-bounds}). 


It is often argued, however, that the above cosmological constraint
can be circumvented if the baryon number and 
one of the lepton flavor numbers are sufficiently
conserved in these R-parity violating couplings, because $B/3-L_i$
for each lepton flavor is separately conserved by the sphaleron
process. (Here, $B$ denotes the baryon number and $L_i$ is the lepton
flavor number of the $i$-th generation.) For instance, if $\lambda'_{1jk}
\lsim 10^{-7}$ is satisfied, then $\lambda'_{2jk}$ and $\lambda'_{3jk}$ can be
much larger than $10^{-7}$. 


However, lepton flavor is not conserved in generic SUSY models
because of the mixings in the slepton mass matrices. 
These slepton mixings can then
erase the asymmetry between the lepton flavors, $L_i-L_j$, and hence
$(B-L_i/3)-(B-L_j/3)$~\cite{Davidson}.
In this paper, we investigate 
the effects of the lepton flavor violation (LFV) 
on the cosmological $B-L$ conservation.
We show that, 
if there exist slepton mixings as large as $\theta_{12}\gsim 10^{-4}$ 
and $\theta_{23(13)}\gsim 10^{-5}$, they would erase the
lepton flavor asymmetry $L_i-L_j$ and spoil the separate $B-L_i/3$
conservation. It means that {\it all} the R-parity violating
couplings must satisfy the cosmological constraint.
In particular, this is the case if LFVs
are observed at the current and future experiments such as MEG~\cite{MEG}
and super B-factories~\cite{Akeroyd:2004mj}. 
We also reinvestigate the cosmological bounds on the R-parity violating 
couplings by solving the Boltzmann equations, and show
that the trilinear couplings $\lambda$, $\lambda'$, and $\lambda''$
must be smaller than ${\cal O}(10^{-6}-10^{-7})$,
and the bilinear coupling should satisfy
$\mu_i/\mu\lsim {\cal O}(10^{-6})$.
Finally, we discuss the implications of the bounds on the R-parity violation
for the LHC phenomenology.

\section{Lepton flavor asymmetry in the early universe}

\subsection{Slepton mixings and lepton flavor violation in the early universe}
\label{sec:LFVinCosmo}

In this section we discuss how fast the lepton flavor is mixed under 
LFV processes. Here, we do not introduce any R-parity violations to
concentrate on the effect of the slepton mixings.

Let us first consider the basis where the lepton Yukawa matrix is diagonal, 
which is familiar in the context of low energy LFV phenomenology.
In this basis, we have a term
\beq
W = h_i L_i \bar{E}_i H_d
\eeq
in the superpotential and soft slepton masses
\beq
-{\cal L} = 
(m_{\tilde{l}_L}^2)_{ij} \tilde{L}_i^* \tilde{L}_j
+
(m_{\tilde{e}_R}^2)_{ij} \tilde{\bar{E}}_i^* \tilde{\bar{E}}_j
\eeq
in the Lagrangian.
To discuss the LFV effects in the early universe, however, 
it is more appropriate to take a basis where the slepton mass matrices are
diagonal~\cite{Davidson}. Because the gaugino--slepton--lepton interactions
are stronger than the Yukawa interactions, one should also diagonalize the 
gaugino interactions. Namely, leptons and sleptons are rotated by the 
same unitary matrices which diagonalize the slepton mass matrices.
Assuming that the mixing angles are small,
these rotations are expressed as
\bea
L_i &=& (U_{\tilde{l}_L})_{ij} \hat{L}_j
\;\simeq \;
\hat{L}_i + \sum_{j\ne i}\theta_{ij}^L \hat{L}_j\,,\\
\bar{E}_i &=& (U_{\tilde{e}_R})_{ij} \hat{\bar{E}}_j
\;\simeq \;
\hat{\bar{E}}_i + \sum_{j\ne i}\theta_{ij}^R \hat{\bar{E}}_j\,,
\eea
where $\theta^{L/R}_{ij}\simeq -\theta^{L/R}_{ji}$ are the mixing angles,
and the hat denotes the new basis. Note that those mixing angles
are different from the dimensionless parameters
\beq
(\delta_{\tilde{l}_L})_{ij} = (m_{\tilde{l}_L}^2)_{ij}/(m_{\tilde{l}_L}^2)_{ii}\,,
\quad
(\delta_{\tilde{e}_R})_{ij} = (m_{\tilde{e}_R}^2)_{ij}/(m_{\tilde{e}_R}^2)_{ii}\,,
\eeq
which are familiar in the context of the LFV rare processes.
In fact, the mixing angles $\theta^{L/R}_{ij}$ are enhanced compared to
$(\delta_{\tilde{l}_L / \tilde{e}_R})_{ij}$.
(See discussion in Sec.~\ref{sec:LFV}.)

In this new basis, the LFV effects appear
only in the Yukawa couplings, which are given by
\beq
W_{\rm LFV} 
=
\sum_{i\ne j} h_{ij} 
\hat{L}_i \hat{\bar{E}}_j H_d\,,
\eeq
where $h_{ij} \equiv h_i \theta^R_{ij} + h_j \theta^L_{ji}$\,.
For instance,
\bea
h_{23} &=& h_2 \theta_{23}^R + h_3 \theta_{32}^L
\\
&\simeq &
\left(0.0061\cdot \theta_{23}^R + 0.10\cdot \theta_{32}^L \right)
\lrf{\tan\beta}{10}.
\eea

We now estimate how much the lepton flavor asymmetry $L_i-L_j$ is 
erased due to the above LFV interactions.
To this end, we solve the Boltzmann equation for the evolution of $L_i-L_j$.
Here, for simplicity, we include only the effect of the higgsino decay
and its inverse process, $\tilde{H}\rightleftarrows \tilde{L}_i \bar{E}_j$
and $\tilde{H}\rightleftarrows L_i \tilde{\bar{E}}_j$,
assuming that the higgsino is heavier than the sleptons.
Other processes such as $2\to 2$ scatterings and those with
Higgs bosons may be comparably important, 
but it is expected that the bounds on the mixing angles
will change only by order one factors.
Note that these additional effects only strengthen the erasure effect,
and therefore the bounds we will derive should be regarded
as conservative ones.

As is derived in Appendix \ref{app:LFV}, the Boltzmann equation for the evolution of 
$L_i-L_j$ is given by [see \eqref{eq:BoltzmannN2N3}]
\beq
T\diff{}{T}N_{L_i-L_j}
= 
\frac{16(\Gamma_{ij}+\Gamma_{ji})}{3H}
\frac{\funcTs1{\higgsinomass}}{\funcTs2{\sleptonmass}+2}
N_{L_i-L_j}
\label{eq:BoltzmannNiNj}
\eeq
where $T$ is the temperature of the universe,
$H$ is the Hubble parameter,
$F_i(x)=x^2 K_i(x)$ with $K_i(x)$ being the 
modified Bessel functions of the second kind.
$N_{L_i-L_j}$ is defined as
$N_{L_i-L_j} = 
(n_{L_i}-n_{\bar L_i})/T^3
-(n_{L_j}-n_{\bar L_j})/T^3$,
where $n_{L_i}$ and $n_{\bar L_i}$ denotes the 
lepton and anti-lepton number density
in the $i$-th generation, respectively.
The partial rate $\Gamma_{ij}$ is given by
\beq
\Gamma_{ij}=\frac{|h_{ij}|^2}{32\pi}
m_{\tilde{H}}
\left(
1-\frac{{\sleptonmass}^2}{{\higgsinomass}^2}
\right)^2\,,
\eeq
where 
$\higgsinomass$ and $\sleptonmass$ are the masses of higgsino and sleptons, respectively.
We assume that the slepton masses are approximately the same.
Note that the Boltzmann equation \eqref{eq:BoltzmannNiNj} is symmetric under
the exchange of the left-handed and right-handed slepton mixings,
$\theta_{ij}^L \leftrightarrow \theta_{ij}^R$, i.e.,
they give the same effect on the evolution of $N_{L_i-L_j}$.

\begin{figure}[t]
\begin{center}
\includegraphics[width=8cm]{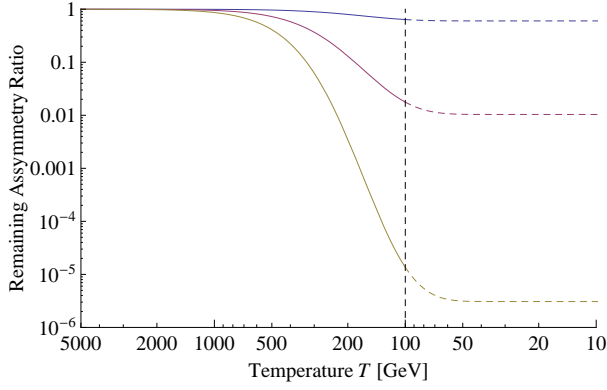}
\end{center}
\caption{Time evolution of $N_{L_2-L_3}$ for slepton mixing angles
$\theta_{23}^{L/R}= 1\times10^{-6}$, $3\times10^{-6}$, and $5\times10^{-6}$,
from the top to the bottom, for
$\higgsinomass = 600\gev$,
$\sleptonmass = 200\gev$, and $\tan\beta=10$.
The vertical dashed line denotes the sphaleron
 decoupling temperature $T_*\simeq 100\gev$.
The normalization is arbitrary.
The time evolution of $N_{L_1-L_3}$ for 
$\theta_{13}^{L/R}= (1-5)\times10^{-6}$ is almost the same.
}
\label{fig:TimeEvolution}
\end{figure}
In Fig.~\ref{fig:TimeEvolution}, the time evolution of $N_{L_2-L_3}$ is shown
for $\theta_{23}^{L/R}\simeq (1-5)\times 10^{-6}$, for $\higgsinomass = 600$ GeV, 
$\sleptonmass = 200$ GeV, and $\tan\beta=10$.
One can see that the flavor asymmetry is rapidly decreased for $T\lsim \higgsinomass$, and
almost washed out
for $\theta_{23}^{L/R}\gsim 3\times 10^{-6}$.
The time evolution of $N_{L_1-L_3}$ 
for $\theta_{13}^{L/R}\simeq (1-5)\times 10^{-6}$ is essentially the same.

\begin{figure}[t]
\begin{center}
\includegraphics[width=8cm]{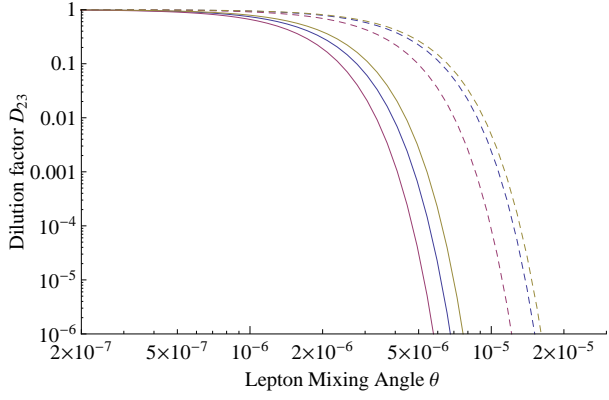}
\end{center}
\caption{
The dilution factor $D_{L_2-L_3}$ ($D_{L_1-L_3}$) as a function of the slepton mixing angle
$\theta_{23}^{L}$ ($\theta_{13}^{L}$) or $\theta_{23}^{R}$ ($\theta_{13}^{R}$),
for $\higgsinomass = 600$, 200 and 1200 GeV, from the left to the right.
The slepton mass $\sleptonmass$ is $0.4\higgsinomass$ for the solid lines and $0.8\higgsinomass$ for the dashed lines.
We took $T_* = 100\gev$ and $\tan\beta=10$.
}
\label{fig:D_i3}
\end{figure}
\begin{figure}[t]
\begin{center}
\includegraphics[width=8cm]{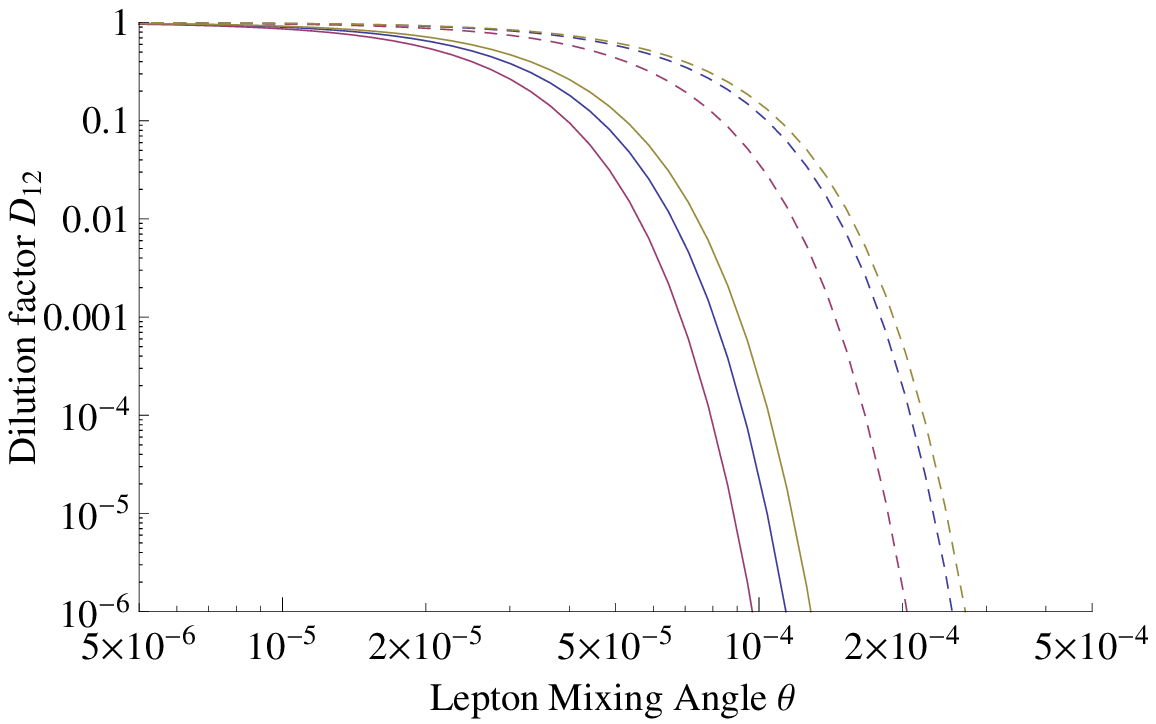}
\end{center}
\caption{
The same as Fig.~\ref{fig:D_i3} but for
the dilution factor $D_{L_1-L_2}$ as a function of the slepton mixing angle
$\theta_{12}^{L/R}$.}
\label{fig:D_12}
\end{figure}

In Fig.~\ref{fig:D_i3} and Fig.~\ref{fig:D_12}, we show the dilution factors 
\beq
D_{L_i-L_j} \equiv \frac{N_{L_i-L_j}(T_*)}{N_{L_i-L_j}(T\gg T_*)}
\eeq
as functions of the mixing angles $\theta_{ij}^{L/R}$, 
where $T_*\sim 100$ GeV is the temperature 
when the sphaleron process is decoupled.
In the numerical calculations, we take $T_* = 100$ GeV, 
$\higgsinomass = 200$, 600 and 1200 GeV,
$\sleptonmass / \higgsinomass = 0.4$ and 0.8,
and $\tan\beta=10$.
Note that the dilution effect is weaker for $\higgsinomass=200\gev$ than for
$600\gev$. 
This is because for $\higgsinomass=200\gev$ the duration of the $L_i-L_j$ erasure is
shorter than for $\higgsinomass=600\gev$.

One can see that the lepton flavor asymmetries $L_2-L_3$, 
$L_1-L_3$, and $L_1-L_2$ are washed away
for
\bea
\theta_{23}^{L/R} &\gsim& (0.3-1.0)\times 10^{-5}\cdot \lrfp{\tan\beta}{10}{-1},
\label{eq:theta23}\\
\theta_{13}^{L/R} &\gsim& (0.3-1.0)\times 10^{-5}\cdot \lrfp{\tan\beta}{10}{-1},
\label{eq:theta13}\\
\theta_{12}^{L/R} &\gsim& (0.6-1.6)\times 10^{-4}\cdot \lrfp{\tan\beta}{10}{-1},
\label{eq:theta12}
\eea
respectively. (We take the value where the dilution factor becomes $D_{L_i-L_j}\simeq 0.01$.)
If any two of these inequalities are simultaneously satisfied,
all lepton flavor numbers become essentially the same, 
$L_1 = L_2 = L_3$, and hence $B-L_1/3 =B-L_2/3 =B-L_3/3$.
As we will see in the next subsection, such slepton mixings are indeed
naturally expected in generic SUSY models.

\subsection{Lepton Flavor Violation}
\label{sec:LFV}

The lepton flavor violation is rather generic in a wide class of the SUSY models. In fact, the 
flavor structures of the neutrinos and quarks can induce mixtures of the slepton generations. 
In the see-saw models, the neutrino Yukawa interaction radiatively contributes to the left-handed 
slepton mass, while the right-handed one receives a correction from the CKM mixings above the 
GUT scale in the SUSY GUT models. By mediating the SUSY breaking effect at the Planck scale 
such as in the gravity mediation, the slepton mass matrices, thus, acquire the flavor mixing through 
the renormalization group evolution down to the weak scale, which are approximately shown as
\begin{eqnarray}
  (m_{\tilde{l}_L}^2)_{ij}  &\simeq&
  - \frac{1}{8\pi^2} 
  (Y_\nu^*)_{ki} (Y_\nu)_{kj}
  (3m_0^2+ a_0^2) 
  \ln\frac{M_P}{M_{R_k}},
  \nonumber\\
  (m_{{\tilde{e}_R}}^2)_{ij}  &\simeq&
  - \frac{3}{8\pi^2} 
  Y_t^2
  (V_{\rm CKM})_{3i} (V_{\rm CKM})^*_{3j}
  (3m_0^2+ a_0^2)
  \ln\frac{M_P}{M_{H_c}},
  \label{eq:RG-mass}
\end{eqnarray}
where the $m_0$ and $a_0$ are typical values of the scalar mass and the trilinear coupling. Here 
$M_P$, $M_{R_k}$ and $M_{H_c}$ are the Planck scale, the mass of the $k$-th right-handed 
neutrino and the mass of the colored Higgs boson of the GUT, respectively. Obviously, the neutrino 
Yukawa couplings, $Y_\nu$, and the CKM matrix, $V_{\rm CKM}$, induce the flavor violation even 
in the absence of the slepton mixings at the input scale. 

The flavor mixings in \eqref{eq:RG-mass} are represented in terms of the parameters at the weak 
scale and those of the right-handed neutrino. It is known that the neutrino mass is expressed as 
$(Y_\nu)_{ij} = 1/\langle H_u\rangle \sqrt{M_{R_i}}R_{ik}\sqrt{m_{\nu_k}}U_{jk}^*$ with the vacuum 
expectation value of the up-type Higgs boson, $\langle H_u\rangle$, the light neutrino mass, 
$m_{\nu_k}$, and the PMNS matrix, $U_{jk}$~\cite{Casas:2001sr}.
Here, $U_{jk}$ and $R_{ij}$ satisfying $R_{ik}R_{jk} 
= \delta_{ij}$ generally have a nontrivial flavor structure. 
On the other hand, we assumed the minimal Yukawa interactions above the GUT scale in the 
latter line of \eqref{eq:RG-mass}. Besides, we took only the top Yukawa coupling, $Y_t$, into the 
account because the up-type quark masses are hierarchical. Noting that the diagonal components 
of the slepton mass matrices are approximated to be $m_0^2$, we estimate the flavor mixings as
\begin{eqnarray}
  (\delta_{\tilde{l}_L})_{ij} &\sim& 10^{-5} \left(\frac{M_R}{\GEV{10}}\right), \nonumber\\
  (\delta_{\tilde{e}_R})_{21} &\simeq& 3 \times 10^{-4},~~
  (\delta_{\tilde{e}_R})_{32} \,\simeq\, 3 \times 10^{-2},~~
  (\delta_{\tilde{e}_R})_{31} \,\simeq\, 7 \times 10^{-3},
  \label{eq:RG-mixing}
\end{eqnarray}
where $(\delta_A)_{ij}$ is defined as $(\delta_A)_{ij} = (m_A^2)_{ij}/(m_A^2)_{ii}$ $(i \neq j)$,
and we have assumed $a_0^2 = m_0^2$ and $M_{H_c}\simeq 2\times 10^{16}\gev$. 
We also chose the typical mass of the light neutrino to be 0.1eV and 
$R_{ij} = {\cal O}(1)$ discarding cancellations. 

We find that the typical size of the flavor violation is large enough to erase the flavor dependence 
of $B/3-L_i$. Namely, we cannot evade the wash-out condition by postulating the flavorful R-parity 
violation. This conclusion may be seen simply by equating $\theta^{L/R}_{ij}$ in 
Eqs.~(\ref{eq:theta23})--(\ref{eq:theta12}) with $(\delta_{\tilde{l}_L/\tilde{e}_R})_{ij}$ 
in \eqref{eq:RG-mixing}.
Actually, the former becomes much larger than the latter when the 
slepton (sneutrino) is almost degenerate. They have a relation as
\beq
\theta_{ij}^L \simeq
\lrf{m_{\tilde{l}_L}^2}{\Delta m_{\tilde{l}_L}^2} 
(\delta_{\tilde{l}_L})_{ij}\,,
\qquad
\theta_{ij}^R \simeq
\lrf{m_{\tilde{e}_R}^2}{\Delta m_{\tilde{e}_R}^2} 
(\delta_{\tilde{e}_R})_{ij}\,,
\label{eq:theta_delta}
\eeq
where $\Delta m_A^2$ is the difference of the slepton 
(sneutrino) mass eigenvalues, while 
$m_{\tilde{l}_L}^2$ and $m_{\tilde{e}_R}^2$ in the numerators denote the average of
the mass eigenvalues.
In many models, the slepton (sneutrino) mass matrices 
are set to be universal at high scale to avoid too large lepton-flavor violations. 
Then, the mass difference stems 
from the Yukawa interactions through the renormalization group evolution. 
Since the effect is very small, $\theta^{L/R}_{ij}$ becomes much larger than 
$(\delta_A)_{ij}$, leading to the rapid mixture of 
the sleptons in the early universe. 

The above flavor mixings in \eqref{eq:RG-mass}
induce lepton-flavor violating decays of muon and tau-lepton. The branching 
ratio of the $\mu \to e\gamma$ decay is approximately obtained as~\cite{Hisano:1995cp}\footnote{We neglect the 
contributions 
from multiple flavor mixings such as $(\delta_A)_{23} (\delta_A)_{31}$. }
\begin{equation}
\br(\mu \to e\gamma) \sim
10^{-(12-13)} 
\left(\frac{(\delta_{\tilde{l}_L})_{21}}{10^{-4}}\right)^2 
\left(\frac{\tan\beta}{10}\right)^2
\left(\frac{m\s{soft}}{400\gev}\right)^{-4},
\label{eq:mu-egamma}
\end{equation}
where $m\s{soft}$ is a typical mass of the sleptons, sneutrinos, neutralinos and charginos. Here, 
the coefficient is quite sensitive to the details of the sparticle mass spectrum. In the presence 
of the right-handed slepton mixing, $(\delta_{\tilde{e}_R})_{21}$, we checked that the additional 
contribution is smaller by a factor or by an order of magnitude than \eqref{eq:mu-egamma}. 
We notice that the decay rate is proportional to $(\delta_A)_{ij}^2$ rather than 
$(\theta_{ij})^2$. 
Thus, when $\Delta m_A^2$ is small, the rate becomes suppressed for fixed $\theta_{ij}$, as is 
expected from the GIM mechanism. 

In Sec.~\ref{sec:LFVinCosmo}, we found that the flavor mixing between $L_1$ and $L_2$
becomes effective in the early universe 
when $\theta^{L/R}_{12}$ satisfies \eqref{eq:theta12}. {}From \eqref{eq:theta_delta}
and \eqref{eq:mu-egamma}, this condition corresponds to
\begin{equation}
\br(\mu \to e\gamma) \gsim
10^{-(12-13)} 
\left(\frac{\Delta m_{\tilde{l}_L}^2}{m_{\tilde{l}_L}^2}\right)^2
\left(\frac{m\s{soft}}{400\gev}\right)^{-4},
\end{equation}
or similar condition with $m_{\tilde{e}_R}^2$.
On the other hand, the current experimental bound and the future sensitivity are
\begin{eqnarray}
\br(\mu \to e\gamma) &<& 1.2 \times 10^{-11}, \nonumber\\
\br(\mu \to e\gamma) &\gsim& 10^{-13},
\end{eqnarray}
from the MEGA~\cite{Brooks:1999pu} and MEG~\cite{MEG} experiments, respectively. 
Comparing these two results, we conclude 
that if we measure $\mu \to e\gamma$ decay in future, 
it suggests that the mixture of the lepton flavor takes 
place effectively in the early universe.

Let us next consider the flavor violating decays of the tau lepton such as $\tau \to \mu(e)\gamma$. 
The branching ratios are estimated as~\cite{Hisano:1995cp} 
\begin{equation}
\br(\tau \to \mu(e)\gamma) \sim
10^{-(13-14)} 
\left(\frac{(\delta_{\tilde{l}_L})_{32(1)}}{10^{-4}}\right)^2 
\left(\frac{\tan\beta}{10}\right)^2
\left(\frac{m\s{soft}}{400\gev}\right)^{-4}.
\end{equation}
The right-handed slepton mixing contributes again by a factor or an order of magnitude smaller 
than the left-handed one. On the other hand, the current experimental bounds are
\begin{eqnarray}
\br(\tau\to\mu\gamma) &<& 4.5 \times 10^{-8}, \nonumber\\
\br(\tau\to e \gamma)   &<& 1.1 \times 10^{-7},
\end{eqnarray}
from Belle and BaBar~\cite{PDG}.
 The sensitivity of the branching ratio is 
expected to be improved by an order of magnitude in future super
B-factories~\cite{Akeroyd:2004mj}. 
Compared with the results of Eqs.~(\ref{eq:theta23}) and (\ref{eq:theta13}), 
we find that the cosmological 
flavor mixing of the tau lepton becomes effective when the branching ratios satisfy
\begin{equation}
\br(\tau \to \mu(e)\gamma) \gsim
10^{-(15-16)} 
\left(\frac{\Delta m_{\tilde{l}_L}^2}{m_{\tilde{l}_L}^2}\right)^2
\left(\frac{m\s{soft}}{400\gev}\right)^{-4}.
\end{equation} 
This lower bound is much smaller than the future sensitivity.
Therefore, if we observe the tau-lepton flavor violation in future colliders, 
then the conditions Eqs.~(\ref{eq:theta23}) and/or (\ref{eq:theta13}) are satisfied, 
and the tau-lepton flavor number is not independently conserved in the early universe.


\section{Implications for the R-parity violation}
As we have shown in Sec.~\ref{sec:LFVinCosmo}, 
if the slepton mixing angles satisfy 
at least two of Eqs.~(\ref{eq:theta23})--(\ref{eq:theta12}), 
then all the lepton flavor asymmetries are equilibrated, 
i.e., $L_1 = L_2 = L_3$. 
In fact, as was discussed in Sec.~\ref{sec:LFV},
sizable slepton mixings are expected in a wide class 
of SUSY models, which are sufficiently large to satisfy Eqs.~(\ref{eq:theta23})--(\ref{eq:theta12}).
In this section, we discuss the bounds on the R-parity violating couplings
in the presence of such lepton flavor violations.

\subsection{Cosmological bounds on the R-parity violation in the presence of slepton mixings}
We assume that at least two of Eqs.~(\ref{eq:theta23})--(\ref{eq:theta12}) are satisfied,
and hence all $B-L_i/3$ are equilibrated.
Then, in order to avoid the baryon erasure,
any of $B-L_i/3$ violating processes should not become effective
before the electroweak transition.

We calculate the dilution factor
\beq
D_{B-L} = \frac{N_{B-L}(T_*)}{N_{B-L}(T\gg T_*)}
\eeq
as functions of the R-parity violating couplings 
$\lambda_{ijk}$, $\lambda'_{ijk}$, $\lambda''_{ijk}$, and $\mu_i$.\footnote{We do not discuss
the bounds on the R-parity violating soft terms, for simplicity.}
The corresponding Boltzmann equations are shown in Appendix \ref{app:RpV}.
The results are shown in Figs.~\ref{fig:UDD}--\ref{fig:LHu}.
Here, for simplicity, we have assumed that all sleptons and all squarks have the same masses $\sleptonmass$
and $\squarkmass$, respectively.

{}From the figures, one can see that the couplings should satisfy
\bea
\sqrt{\sum_{ijk}|\lambda''_{ijk}|^2} &\lsim& (4-5)\times 10^{-7}\,, \label{eq:boundUDD} \\
\sqrt{\sum_{ijk}|\lambda'_{ijk}|^2} &\lsim& (3-6)\times 10^{-7}\,, \\
\sqrt{\sum_{ijk}|\lambda_{ijk}|^2} &\lsim& (0.6-1)\times 10^{-6}\,, \label{eq:boundLLE}\\
\sqrt{\sum_{i}\left|\frac{\mu_i}{\mu} \right|^2} &\lsim& (1-2)\times 10^{-6}
\lrfp{\tan\beta}{10}{-1}\,, \label{eq:boundLHu}
\eea
for $\squarkmass \simeq 200 - 1200\gev$ and $\sleptonmass \simeq 100 - 400\gev$.
(Again, we took the value where the dilution of the $B-L$ becomes $D_{B-L}\simeq 0.01$.)
We should note that the bound on the $\UDD$ coupling $\lambda''_{ijk}$ in \eqref{eq:boundUDD}
applies even without the lepton flavor violation.

\begin{figure}[t]
\begin{center}
\includegraphics[width=8cm]{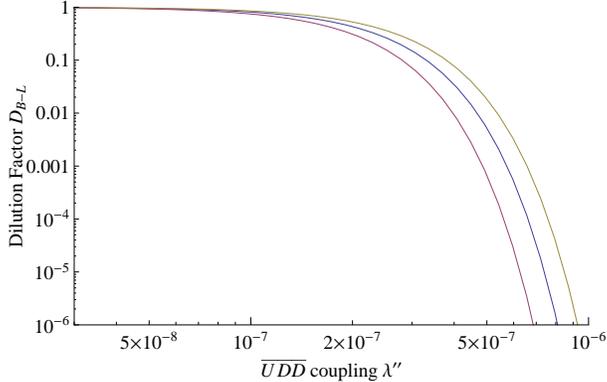}
\end{center}
\caption{
The dilution factor $D_{B-L}$ in the presence of an R-parity violating term
$\lambda''\bar U_i\bar U_j \bar D_k$
for $\squarkmass = 600$, 200 and 1200 GeV, from the left to the right.
We took $\sleptonmass=100$GeV and $\higgsinomass=300$GeV, but this result is nearly independent of these masses. Other parameters are:  $\tan\beta=10$ and $T_*=100$GeV.
}
\label{fig:UDD}
\end{figure}
\begin{figure}[t]
\begin{center}
\includegraphics[width=8cm]{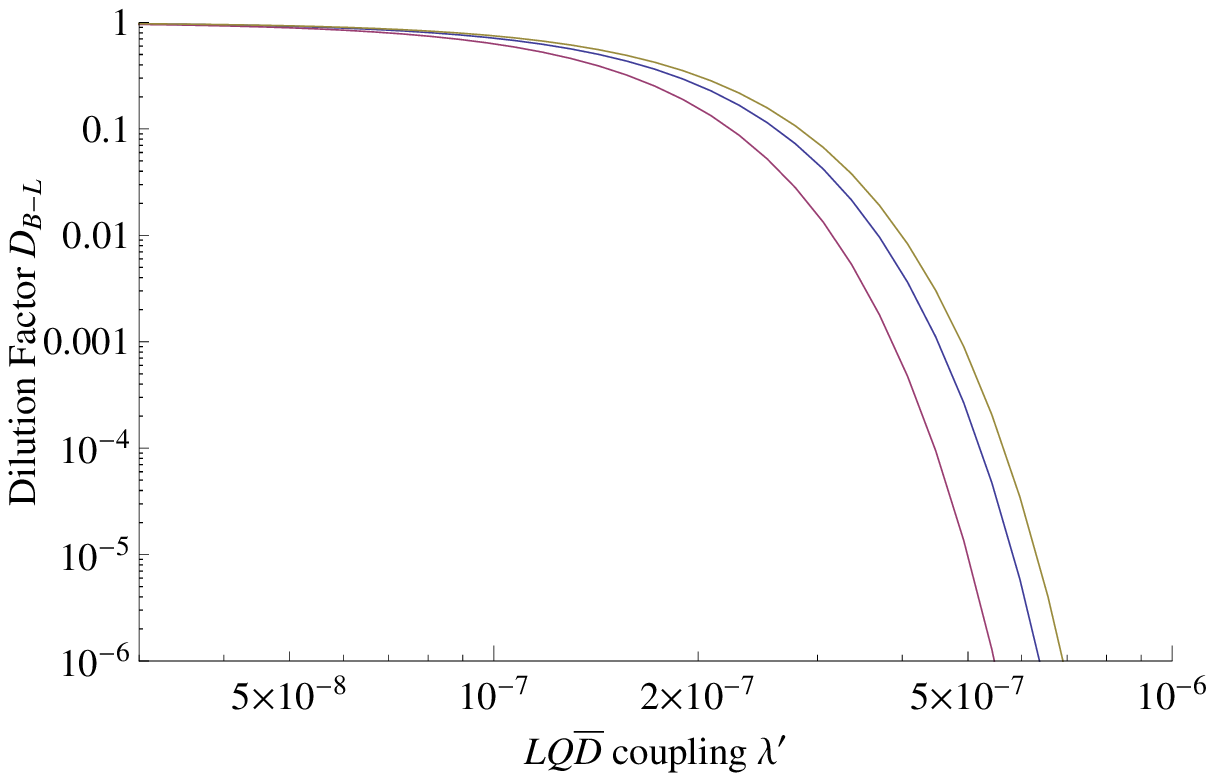}
\end{center}
\caption{
The same as Fig.~\ref{fig:UDD} for $\lambda' L_i Q_j \bar D_k$ interaction.
Parameters are the same as Fig.~\ref{fig:UDD}. $\sleptonmass$ and $\higgsinomass$ hardly affect the result again.
}
\label{fig:LQD}
\end{figure}
\begin{figure}[t]
\begin{center}
\includegraphics[width=8cm]{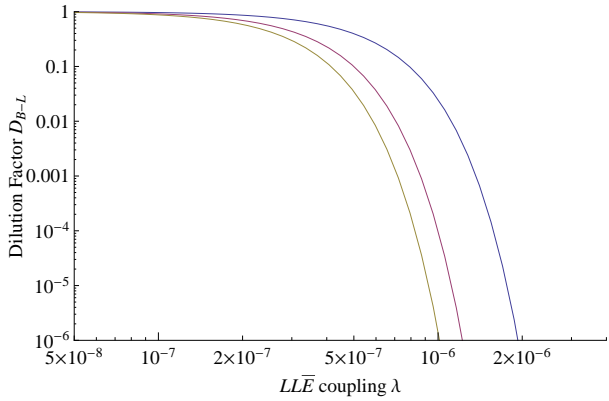}
\end{center}
\caption{
The same as Fig.~\ref{fig:UDD} for $\lambda L_i L_j \bar E_k$ interaction.
$\sleptonmass = 400$, 200 and 100 GeV, from the left to the right, $\squarkmass=600$GeV and $\higgsinomass=300$GeV. In this case the result depends on $\sleptonmass$, and is almost independent of the other masses.
}
\label{fig:LLE}
\end{figure}

\begin{figure}[t]
\begin{center}
\includegraphics[width=8cm]{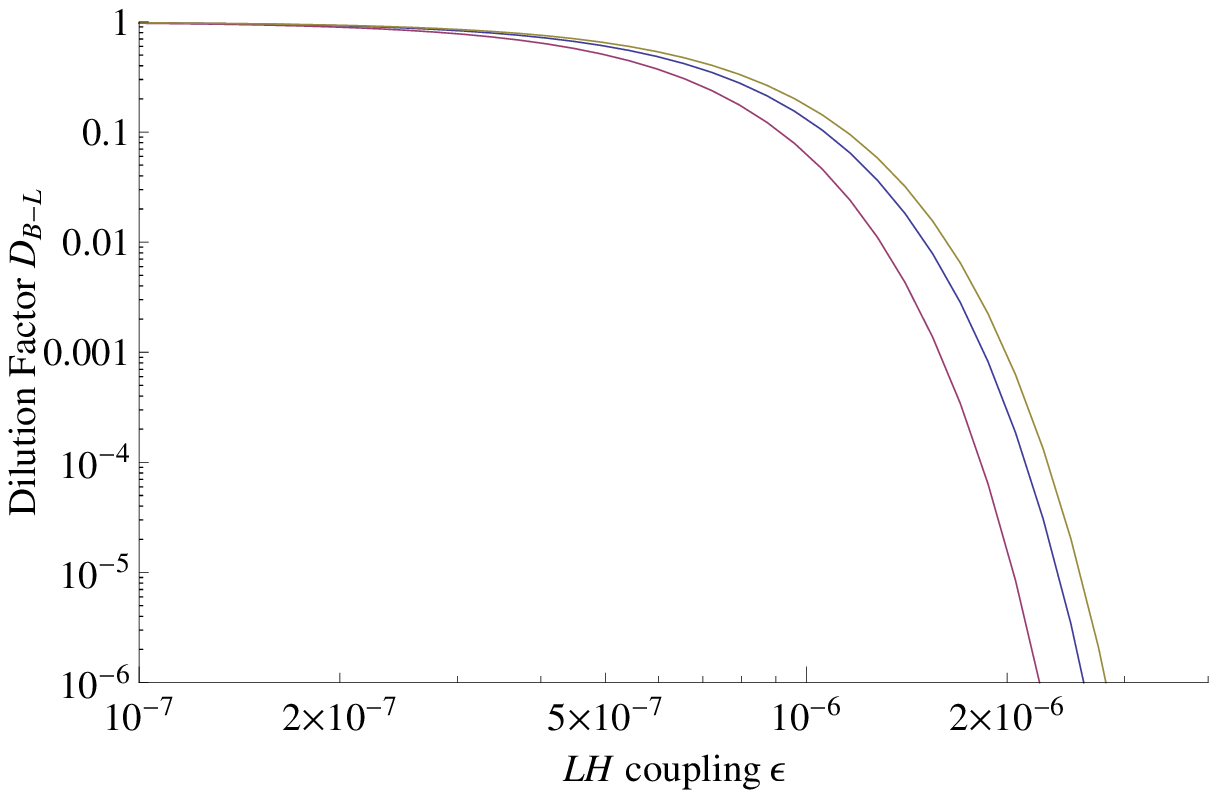}
\end{center}
\caption{
The same as Fig.~\ref{fig:UDD} in the presence of a bilinear R-parity violating
term $\mu_i L_i H_u$ as a function of $\epsilon_i \equiv \mu_i/\mu$. 
The masses and the other parameters are the same as Fig.~\ref{fig:UDD}. 
The result is nearly independent of $\sleptonmass$, $\higgsinomass$, and 
the generation index $i$.
}
\label{fig:LHu}
\end{figure}

\subsection{Implications for collider phenomenology}
In the presence of slepton mixings, all the R-parity violating couplings must satisfy
Eqs.~(\ref{eq:boundUDD})--(\ref{eq:boundLHu}) in order to avoid the baryon erasure.
Interestingly, this means that the LOSP has a long decay length at the LHC. 
For instance, suppose that the LOSP is the stau, mainly consisting of the 
right-handed stau. If the $LL\bar{E}$ coupling $\lambda_{ij3}$ saturates the cosmological bound
\eqref{eq:boundLLE}, the decay length of the stau becomes 
\beq
c\tau_{\tilde \tau} \;\simeq\; 50~{\rm \mu m} 
\lrfp{\lambda_{ij3}}{10^{-6}}{-2}
\lrfp{m_{\tilde \tau}}{100\gev}{-1}\,.
\eeq
This is comparable to the tau-lepton decay length ($c\tau_\tau\simeq 87~{\rm \mu m}$),
which can be probed at the LHC.
Note that this is the shortest possible decay length, and in general 
a (much) longer decay length is expected. 
If the dominant decay of the stau is caused by 
$\lambda_{ijk}$ ($k\ne 3$) or the $LQ\bar{D}$ coupling $\lambda'_{ijk}$, 
the decay length becomes longer since the decay rate is suppressed 
by the left-right mixing of the stau and/or the flavor mixing.

Similar results are obtained for other LOSP cases.
Depending on what is the LOSP and the pattern of the R-parity breaking, 
the dominant decay mode of the LOSP can become three- 
or four-body decay~\cite{Barbier:2004ez}, 
which makes the decay length even longer.
It is important to study the LHC phenomenology of R-parity violating SUSY models
under the cosmological bounds Eqs.~(\ref{eq:boundUDD})--(\ref{eq:boundLHu})
for different LOSP candidates and the different pattern of R-parity violating couplings,
and we leave it for future work.

\section{Conclusion and discussions}

In this paper, implications of the slepton flavor mixings for the 
cosmological constraints on the R-parity violating couplings are discussed.
We have shown that tiny slepton mixing angles 
$\theta_{12}\gsim {\cal O}(10^{-4})$ 
and $\theta_{23}, \theta_{13}\gsim {\cal O}(10^{-5})$ will spoil the 
separate $B/3-L_i$ conservation.
Such slepton mixings are indeed expected in generic SUSY models.
On the other hand, if lepton flavor violations are observed 
in experiments such as MEG and B-factories, it will imply that 
{\it all} the R-parity violating couplings must be suppressed
to avoid the $B-L$ erasure.

We also reinvestigated the cosmological constraints on the R-parity violating couplings 
in the presence of lepton flavor mixings in the slepton sector, 
and showed that the R-parity violating couplings must satisfy 
Eqs.~(\ref{eq:boundUDD})--(\ref{eq:boundLHu}).
It then suggests that the LSP (or NLSP in the gravitino LSP scenario) has 
a long decay length, which can be probed at the LHC.

Interestingly, for such a small R-parity violation,
the gravitino becomes an attractive candidate for the dark matter~\cite{TY,BCHIY}, 
where the gravitino lifetime
can be much longer than the age of the universe due to the double
suppression by the inverse Planck mass and the R-parity breaking
coupling.

So far in this paper, we have assumed that the baryon asymmetry was generated before the electroweak transition,
and found that if LFV events are observed, then the R-parity violation must be so small that the
LSP becomes long-lived in the LHC. This argument can be put the other way around.
If LFVs are discovered and also a sizable R-parity violation is observed at the LHC, 
then it means that the baryon asymmetry of the universe must have been generated 
{\it after} the electroweak transition, as in the electroweak baryogenesis 
and in Affleck-Dine baryogenesis with a long lived condensate or Q-ball.
This is also quite an exciting possibility.

\section*{Acknowledgment}
We thank S.~Matsumoto, K.~Nakaji and T.~Yamashita for useful discussions,
and the YITP workshop ``Summer Institute 2009 on Particle Physics Phenomenology"
 (YITP--W--09--08), August 13--20, 2009, where this work has been initiated. 
The work of K.H. was supported by JSPS Grant-in-Aid for Young Scientists (B) (21740164).
This work was supported by World Premier International Center Initiative (WPI Program),
MEXT, Japan.

\appendix

\section{Boltzmann equations for lepton flavor and R-parity violating processes}
In this Appendix we derive the Boltzmann equations for lepton flavor and R-parity violating processes in the MSSM.

\subsection{Boltzmann equation}
The time evolution of the number density $n_A$ of certain particle $A$ obeys the Boltzmann equation.
When we consider only the part of the time evolution induced by a process $X\rightleftarrows AY$,
that is, the decay of some particle $X$ and its inverse process,
the Boltzmann equation is
\begin{equation}
\restrict{\diff{}{t}n_A + 3Hn_A}{X\rightleftarrows AY} =
n_X\bigvev{\Gamma_{X\to AY}}-n_An_Y\bigvev{(\sigma v)_{AY\to X}},
\label{eq:Boltzmann_in_n}
\end{equation}
where $H$ is the Hubble parameter, and $\vev\ $ denotes thermal average. We assume that $X$, $A$, and $Y$ are all in thermal bath, and discuss the effect of (very weak) $X\rightleftarrows AY$ process. Using the ``yield'' $N=n/T^3$ as a variable, \eqref{eq:Boltzmann_in_n} becomes\footnote{We have used $dT/dt = -HT$, assuming for simplicity that the effective degrees of freedom $g_{*s}(T)$ is constant.}
\begin{equation}
\restrict{T \diff{}T N_A}{X\rightleftarrows AY} =
-\frac1{HT^3} \Bigl[
n_X\bigvev{\Gamma_{X\to AY}}
- n_An_Y\bigvev{(\sigma v)_{AY\to X}}
\Bigr].\label{eq:Boltzmann_in_Y}
\end{equation}

For simplicity, we approximate the distributions of the particles as
the Maxwell-Boltzmann distributions.\footnote{Even if one uses the Bose-Einstein and the Fermi-Dirac distributions, the following discussion is almost unchanged. In this case, one has to include the statistical factors for the final states, and the functions $F_i(x)$ are replaced with modified functions.} Then, the number density $n$ of a particle is
\begin{align}
n
= g \int\dP \exp\left(-\frac{E-\mu}T\right)
  = \frac{g}{2\pi^2}T^3\funcT2{m}\exp\left(\frac{\mu}T\right)
\label{eq:n_in_m_and_mu}
\end{align}
where $\mu$, $g$ and $m$ are the chemical potential, the degree of freedom,
and the mass of the particle, respectively, and $F_i(x)$ is defined as
\begin{equation}
F_i(x)\equiv x^2K_i(x)
\end{equation}
through the modified Bessel function $K_i(x)$ of the second kind.
For a massless particle, $F_2(0) = 2$.
The rate of the decay process is
\begin{align}
n_X\bigvev{\Gamma_{X\to AY}}
&= g_X \int\dP[p_X] \exp\left(-\frac{E_X-\mu_X}T\right)\frac{m_X}{E_X}\Gamma^0_{X\to AY}\\
&= g_X \Gamma^0_{X\to AY}\frac{T^3}{2\pi^2}\funcT1{m_X}
      \exp\left(\frac{\mu_X}T\right).
\end{align}
Here $\Gamma^0_{X\to AY}$ is the partial decay rate in $X$'s rest frame.
Similarly one can calculate the rate of the inverse decay process, which is
\begin{align}
n_An_Y\bigvev{(\sigma v)_{AY\to X}}
&= g_X \Gamma^0_{X\to AY}
 \frac{T^3}{2\pi^2}\funcT1{m_X}
 \exp\left(\frac{\mu_A+\mu_Y}T\right).
\end{align}
Thus when the process $X\rightleftarrows AY$ is in thermal equilibrium,
which means $\mu_X=\mu_A+\mu_Y$,
the decay process and inverse decay process occurs with the same rate.

Here we introduce the difference between decay and inverse decay process, or ``effective decay rate,'' as follows:
\begin{align}
\effvev{X\rightleftarrows AY}
&\equiv  n_X\bigvev{\Gamma_{X\to AY}} - n_An_Y\bigvev{(\sigma v)_{AY\to X}}\\
&=  g_X \Gamma^0_{X\to AY}
  \frac{T^3}{2\pi^2}\funcT1{m_X}
  \left[\exp\left(\frac{\mu_X}T\right)-\exp\left(\frac{\mu_A+\mu_Y}T\right)\right].
\end{align}
The equation~\eqref{eq:Boltzmann_in_Y} is now
\begin{equation}
\restrict{T \diff{}{T}N_A}{X\rightleftarrows AY}
=-\frac{1}{HT^3}\effvev{X \rightleftarrows AY}.
\end{equation}
Moreover, as the chemical potential $\mu_{\bar A}$ of the antiparticle is equal to $-\mu_A$,
the effective rate of the processes of antiparticles are
\begin{equation}
\effvev{\bar X\rightleftarrows \bar A\,\bar Y} 
=  g_X \Gamma^0_{X\to AY}
  \frac{T^3}{2\pi^2}\funcT1{m_X}
  \left[\exp\left(\frac{-\mu_X}T\right)-\exp\left(-\frac{\mu_A+\mu_Y}T\right)\right],
\end{equation}
and therefore
\begin{align}
\restrict{T \diff{}{T}\bigl(N_A-N_{\bar A}\bigr)}{X\rightleftarrows AY}
&= -\frac{1}{HT^3} \left[\effvev{X\rightleftarrows AY} - \effvev{\bar X\rightleftarrows \bar A\,\bar Y}\right]\\
&= -\frac{g_X}{\pi^2}
\frac{\Gamma^0_{X\to AY}}{H}\funcT1{m_X}
  \left[\sinh\left(\frac{\mu_X}T\right)-\sinh\left(\frac{\mu_A+\mu_Y}T\right)\right].
\end{align}

\subsection{Lepton flavor violation}
\label{app:LFV}
Here, we derive the Boltzmann equation for the LFV process in the early universe.
As an example, we consider as LFV processes those induced by the following term 
in the superpotential:
\beq
W = h_{23} L_2\bar E_3\Hd\,.
\eeq
For simplicity, we discuss only the decays and inverse decays of the higgsinos $\tilde{H}$,
and assume that all sleptons have the same mass $\sleptonmass$ ($< \higgsinomass$).

We define the asymmetry $N_{[A]}$ of a supermultiplet as, e.g. for $\muL$,
\begin{equation}
N_{[\muL]}\equiv    \bigl( N_{\muL} - N_{\bar{\muL}}\bigr)
           + \bigl( N_{\tilde\muL} - N_{\tilde\muL^*} \bigr).
\end{equation}
Since leptons are massless before the electroweak transition, the asymmetry is
\begin{align}
N_{[\muL]}&=
\frac{g_{\tilde{\muL}}}{2\pi^2}\funcT2{\sleptonmass}
\left[\exp\left(\frac{\mu_{L_2}}T\right)-\exp\left(\frac{-\mu_{L_2}}T\right)\right]
+ \frac{2g_{\muL}}{2\pi^2}
\left[\exp\left(\frac{\mu_{L_2}}T\right)-\exp\left(\frac{-\mu_{L_2}}T\right)\right]\notag\\
&=
\frac{1}{\pi^2}
\left[\funcT2{\sleptonmass}+ 2\right]
\sinh\left(\frac{\mu_{L_2}}T\right),
\label{eq:WholeYieldOfSuperfield}
\end{align}
and its time evolution induced by LFV processes is described as
\begin{align}
\restrict{T \diff{}{T}N_{[\muL]}}{\rm LFV} &=
-\frac{1}{HT^3}
\biggl[
\effvev{{\tilde H^0}\rightleftarrows \muL \tilde\tauR^*}
+ \effvev{{\tilde H^0}\rightleftarrows \tilde\muL \bar{\tauR}}
\notag\biggr.\\&\hspace{96pt}\biggl.
- \text{(their antiparticles' processes)}
\biggr]\\
&= -2\cdot\frac{2}{\pi^2}%
\frac{\Gamma}{H}\funcT1{\higgsinomass}
  \left[\sinh\left(-\frac{\mu_{\Hd}}{T}\right)-\sinh\left(\frac{\mu_{L_2}+\mu_{\bar E_3}}T\right)\right],
\end{align}
where
\beq
\Gamma = \frac{|h_{23}|^2}{32\pi}
m_{\tilde{H}}
\left(
1-\frac{{\sleptonmass}^2}{m_{\tilde{H}}^2}
\right)^2
\eeq
is the partial decay rate of each process, which is the same for all four processes.
Similarly,
\begin{align}
\restrict{T \diff{}{T}N_{[\numu]}}{\rm LFV}
&= -2\cdot\frac{2}{\pi^2}
\frac{\Gamma}{H}\funcT1{\higgsinomass}
  \left[\sinh\left(-\frac{\mu_{\Hd}}{T}\right)-\sinh\left(\frac{\mu_{L_2}+\mu_{\bar E_3}}T\right)\right],\\
\restrict{T \diff{}{T}N_{[\tauR]}}{\rm LFV}
&= -4\cdot\frac{2}{\pi^2}
\frac{\Gamma}{H}\funcT1{\higgsinomass}
  \left[\sinh\left(\frac{\mu_{\Hd}}{T}\right)-\sinh\left(\frac{-\mu_{L_2}-\mu_{\bar E_3}}T\right)\right].
\end{align}

Now let us consider the asymmetry of each lepton flavor, which is defined as
\begin{equation}
N_2 \equiv  N_{[\muL]}+N_{[\muR]}+N_{[\numu]},
\end{equation}
and so on.
{}From \eqref{eq:WholeYieldOfSuperfield}, they are given by
\beq
N_i
\simeq
\frac{1}{\pi^2}
\left[\funcT2{\sleptonmass} + 2\right]
\frac{2\mu_{L_i}-\mu_{\bar E_i}}T.
\label{eq:Ni}
\eeq
Under the LFV interaction $L_2\bar E_3\Hd$, the time evolution of the difference $N_2-N_3$ is
given by
\begin{align}
T\diff{}{T}\bigl(N_2-N_3\bigr)
&= \diff{}{T}\bigl(N_{[\muL]}+N_{[\numu]}-N_{[\tauR]}\bigr)\\
&= \frac{16}{\pi^2}
\frac{\Gamma}{H}\funcT1{\higgsinomass}
  \left[\sinh\left(\frac{\mu_{\Hd}}{T}\right)+\sinh\left(\frac{\mu_{L_2}+\mu_{\bar E_3}}T\right)\right]\\
&\simeq \frac{16}{\pi^2}
\frac{\Gamma}{H}\funcT1{\higgsinomass}
  \left[\frac{\mu_{\Hd}+\mu_{L_2}+\mu_{\bar E_3}}{T}\right],
\end{align}
where we have used $\mu\ll T$.
On the other hand, reactions mediated by the diagonal lepton Yukawa couplings 
are in thermal equilibrium
for $T\lsim \GEV{5}$, which leads to
\beq
\mu_{L_i} + \mu_{\bar E_i} + \mu_{H_d} = 0\,,
\eeq
and hence
\begin{align}
T\diff{}{T}\bigl(N_2-N_3\bigr)
&\simeq 
\frac{16}{\pi^2}
\frac{\Gamma}{H}\funcT1{\higgsinomass}
  \left[\frac{(2\mu_{L_2}-\mu_{\bar E_2})-(2\mu_{L_3}-\mu_{\bar E_3})}{3T}\right].
\end{align}
Therefore, from \eqref{eq:Ni} one obtains
\beq
T\diff{}{T}\bigl(N_2-N_3\bigr)
= 
\frac{16\Gamma}{3H}
\frac{\funcTs1{\higgsinomass}}{\funcTs2{\sleptonmass}+2}
\left(N_2-N_3\right).
\label{eq:BoltzmannN2N3}
\eeq

\subsection{R-parity violation}
\label{app:RpV}
The evolution of the $B-L$ asymmetry under the R-parity violating 
interactions can be discussed in the similar way as Sec.~\ref{app:LFV}.
To this end, we first discuss the relations between the chemical potentials and the $B-L$ asymmetry in Sec.~\ref{sec:chemical}.
We then discuss the time evolution of the $B-L$ asymmetry in the presence of R-parity violating couplings
in Sec.~\ref{sec:UDD}--Sec.~\ref{sec:LHu}.

\subsubsection{$B-L$ asymmetry and chemical potentials}
\label{sec:chemical}
For $\GEV{2}\lsim T \lsim \GEV{5}$, the reactions mediated by Yukawa and gauge interactions as well as the sphaleron process are all in thermal equilibrium, leading to the following relations between the chemical potentials of the MSSM particles
\bea
\mu_{Q} + \mu_{\bar U} &=& -\mu_{H_u} = \mu_{H_d}\,,\\
\mu_{Q} + \mu_{\bar D} &=& - \mu_{H_d}\,,\\
\mu_{L_i} + \mu_{\bar E_i} &=& - \mu_{H_d}\,,\\
9\mu_Q + \sum_i \mu_{L_i} &=& 0\,.
\eea
Note that the chemical potentials of gauge bosons and gauginos vanish, and hence particles in the same super- and gauge multiplet have the same chemical potentials. Moreover, quark mixings make their chemical potentials independent of 
the flavors.
Thus, all the chemical potentials can be expressed in terms of four of them, 
e.g., $\mu_{L_i}$ and $\mu_{H_d}$:
\bea
\mu_Q &=& -\frac{1}{9}\sum_i \mu_{L_i}\label{eq:muQ}\,,\\
\mu_{\bar U} &=& \frac{1}{9}\sum_i \mu_{L_i} + \mu_{H_d}\,,\\
\mu_{\bar D} &=& \frac{1}{9}\sum_i \mu_{L_i} - \mu_{H_d}\,,\\
\mu_{\bar E_i} &=& - \mu_{L_i} - \mu_{H_d}\label{eq:muE}\,.
\eea
Let us define the asymmetry of a supermultiplet $N_{[A]}$ as
\begin{equation}
N_{[A]}\equiv    \bigl( N_{A} - N_{\bar{A}}\bigr)
           + \bigl( N_{\tilde A} - N_{\tilde{A}^*} \bigr).
\end{equation}
Since quarks and leptons are massless before the electroweak transition, 
from \eqref{eq:n_in_m_and_mu}
one obtains
\beq
N_{[A]} =
\frac{g_A}{\pi^2}
\cdot
\funcgeff{m_{\tilde A}}
\sinh\left(\frac{\mu_A}{T}\right)\,,
\quad
g_{\rm eff}(x) \equiv 2 + F_2(x)\,,
\label{eq:WholeYieldOfSuperfield_2}
\eeq
for $A=Q,\bar{U},\bar{D}, L_i, \bar{E_i}$.
Their degrees of freedom are given by
$(g_Q, g_{\bar U}, g_{\bar D}, g_{L_i}, g_{\bar{E_i}})
= (18, 9, 9, 2, 1)$. 
For simplicity, we assume that all the squarks and the sleptons have the same masses
$\squarkmass$ and $\sleptonmass$, respectively. 
Then, the hypercharge conservation
\beq
\sum_A Y_A N_{[A]} = 0
\eeq
leads to
\bea
&&
\funcgeff{\squarkmass}
\left(\frac{3\mu_Q -6\mu_{\bar U}+3\mu_{\bar D}}{T}\right)
+
\funcgeff{\sleptonmass}
\left(\frac{-\sum_i \mu_{L_i} + \sum_i \mu_{E_i}}{T}\right)
\nonumber\\
&+&
\funcgeff{\higgsinomass}
\left(\frac{\mu_{H_u} - \mu_{H_d}}{T}\right)
=0\,,
\eea
where we have used $\mu\ll T$ and neglected the Higgs boson masses for simplicity.
Therefore, 
from Eqs.~(\ref{eq:muQ})--(\ref{eq:muE}) 
we obtain
\beq
\mu_{H_d} = 
-C_{H_d}(T)
\sum_i \mu_{L_i}\,,
\label{eq:HdAndLi}
\eeq
where
\beq
C_{H_d}(T)
=
\frac{1}{3}\cdot
\frac{2\geff(\squarkmass/T)+6\geff(\sleptonmass/T)}
{9\geff(\squarkmass/T)+3\geff(\sleptonmass/T)+2\geff(\higgsinomass/T)}\,.
\label{eq:CHd}
\eeq
Together with Eqs.~(\ref{eq:muQ})--(\ref{eq:muE}),
all the chemical potentials are now given 
in terms of $\mu_{L_i}$.
Note that these relations are independent of the existence of
lepton flavor and R-parity violations.

We define the $B-L$ asymmetry as $N_{B-L}\equiv (n_B-n_L)/T^3$. 
{}From \eqref{eq:WholeYieldOfSuperfield_2},
it is given by, for $\mu\ll T$, 
\bea
N_{B-L} &=&
\frac{1}{3}\left(N_{[Q]}-N_{[\bar U]}-N_{[\bar D]}\right)
-\sum_i \left(N_{[L_i]}-N_{[\bar E_i]}\right)
\nonumber\\
&\simeq& 
-\frac{1}{\pi^2}
C_{B-L}(T)
\frac{\sum_i\mu_{L_i}}T\,,
\label{eq:BLytomu}
\eea
where
\beq
C_{B-L}(T)
=
\frac{4}{3}\funcgeff{\squarkmass}
+3
\Bigl(1-C_{H_d}(T)\Bigr)
\funcgeff{\sleptonmass}
\eeq
and $C_{H_d}(T)$ is given by \eqref{eq:CHd}.
\subsubsection{$\UDD$ interaction}
\label{sec:UDD}
In the presence of the superpotential
\begin{equation}
W = \frac{1}{2}\lambda''_{ijk}\bar{U}_i\bar{D}_j\bar{D}_k,
\end{equation}
the time evolution of $N_{B-L}$ is given by
\bea
T \diff{}{T}\left(N_{B-L}\right)
&\simeq&
- \frac{1}{\pi^2}\cdot
\frac{9\sum_{ijk}\Gamma_{\bar{U}_i\bar{D}_j\bar{D}_k}}{H}
 \funcT1{\squarkmass}
\frac{\mu_{\bar U} + 2\mu_{\bar D}}{T}
\\
&=& 
\frac{1}{H}\sum_{ijk}\Gamma_{\bar{U}_i\bar{D}_j\bar{D}_k}
 \funcT1{\squarkmass}
\frac{3+9\,C_{H_d}(T)}{C_{B-L}(T)}
\cdot
N_{B-L}\,,
\label{eq:BL_UDD}
\eea
where
\begin{equation}
\Gamma_{\bar{U}_i\bar{D}_j\bar{D}_k} =\frac{1}{16\pi} |\lambda''_{ijk}|^2 m_{\tilde{q}}\,.
\end{equation}
We should emphasize that \eqref{eq:BL_UDD} holds even in the absence of 
the lepton flavor violation.

\subsubsection{$L L \bar{E}$ interaction}
\label{sec:LLE}
Here we assume that the lepton flavor asymmetries vanish because of the lepton flavor violation, that is, we use $\mu_{L_1} = \mu_{L_2} = \mu_{L_3}$ in addition to \eqref{eq:BLytomu}.
Under this assumption, the time evolution of $N_{B-L}$ under the superpotential
\begin{equation}
W = \frac{1}{2}\lambda_{ijk} L_i L_j \bar{E}_k
\end{equation}
is described as
\bea
T \diff{}{T}N_{B-L} 
&\simeq &
-
\frac{1}{\pi^2}\frac{3\sum_{ijk}\Gamma_{L_iL_j\bar{E}_k}}{H}
 \funcT1{\sleptonmass}
\frac{2\mu_L + \mu_{\bar E}}{T}
\\
&=&
\frac{1}{H}\sum_{ijk}\Gamma_{L_iL_j\bar{E}_k}
 \funcT1{\sleptonmass}
\frac{1+3\,C_{H_d}(T)}{C_{B-L}(T)} 
\cdot
N_{B-L}\,,
\label{eq:BL_LLE}
\eea
where
\begin{equation}
\Gamma_{L_i L_j \bar{E}_k} = \frac{1}{16\pi} |\lambda_{ijk}|^2 m_{\tilde\ell}.
\end{equation}

\subsubsection{$L Q \bar{D}$ interaction}
\label{sec:LQD}
Here also we assume the vanishment of lepton flavor asymmetries. The time evolution of $N_{B-L}$
under the superpotential
\begin{equation}
W = \lambda'_{ijk} L_i Q_j \bar{D}_k
\end{equation}
is
\begin{align}
 T \diff{}{T}N_{B-L} 
&\simeq
-\frac{1}{\pi^2}
\left[
\frac{12 \sum_{ijk}\Gamma_{\tilde q:L_i Q_j \bar{D}_k}}{H}
\funcT1{\squarkmass}
+
\frac{6\sum_{ijk} \Gamma_{\tilde\ell:L_i Q_j \bar{D}_k}}{H}
\funcT1{\sleptonmass}
\right]
\frac{\mu_L+\mu_Q+\mu_{\bar D}}{T}
\nonumber \\
&=
\frac{1}{H}
\sum_{ijk}
\left[2\Gamma_{\tilde q:L_i Q_j \bar{D}_k}\funcT1{\squarkmass}
+
\Gamma_{\tilde\ell:L_i Q_j \bar{D}_k}\funcT1{\sleptonmass}
\right]
\frac{2+6\,C_{H_d}(T)}{C_{B-L}(T)}
N_{B-L}\,,
\label{eq:BL_LQD}
\end{align}
where
\begin{align}
\Gamma_{\tilde q: L_iQ_j\bar{D}_k} &= \frac1{16\pi} |\lambda'_{ijk}|^2 m_{\tilde q},&
\Gamma_{\tilde \ell: L_iQ_j\bar{D}_k} &= \frac1{16\pi} |\lambda'_{ijk}|^2 m_{\tilde \ell}.
\end{align}

\subsubsection{Bilinear R-parity violation}
\label{sec:LHu}
The bilinear R-parity violating term $\mu_i L_i H_u$ induces, through the $L_i$-$H_d$ mixings, 
effective trilinear couplings $\lambda_{ijk}$ and $\lambda'_{ijk}$.
Then, the time evolution of $B-L$ can be discussed by using the Boltzmann equations in
Sec.~\ref{sec:LLE} and Sec.~\ref{sec:LQD}.

\end{document}